\documentclass[prd,twocolumn,superscriptaddress,nofootinbib,aps,10pt,preprintnumbers]{revtex4-1}

\usepackage[colorlinks=true,linkcolor=black,citecolor=blue,urlcolor=blue, pdfborder={0 0 0}]{hyperref}
\usepackage{mathtools}
\usepackage[utf8]{inputenc}
\usepackage{color}
\usepackage[table,xcdraw,dvipsnames]{xcolor}
\usepackage{subfigure}
\usepackage{amssymb}
\usepackage[normalem]{ulem}

\def\eqs#1#2{{Eqs.~(\ref{#1})--(\ref{#2})}}

\def\app#1{{Appendix~\ref{#1}}}

 \newcommand{\vev}[1]{\langle #1 \rangle}

\renewcommand{\bar}{\overline}

\newcommand{\dchi}{\Delta_\mathcal{X}}
\newcommand{\mX}{\mathcal{X}}
\newcommand{\mXb}{\boldsymbol{\mathcal{X}}}

\newcommand{\X}{{\cal X}}

\newcommand{\U}{{\rm U}}

\newcommand{\gaN}{g_{aN}}
\newcommand{\gan}{g_{an}}
\newcommand{\gap}{g_{ap}}
\newcommand{\gae}{g_{ae}}
\newcommand{\gag}{g_{a\gamma}}

\newcommand{\NDW}{N_{\mathrm{DW}}} 

\newcommand{\beq}{\begin{equation}}
\newcommand{\eeq}{\end{equation}}
\newcommand{\bea}{\begin{eqnarray}}
\newcommand{\eea}{\end{eqnarray}}

\renewcommand{\(}{\left(}
\renewcommand{\)}{\right)}

\newcommand{\eqn}[1]{Eq.~(\ref{#1})}
\newcommand{\eqns}[2]{Eqs.~(\ref{#1})--(\ref{#2})}
\newcommand{\Eqn}[1]{Eq.~(\ref{#1})}

\renewcommand{\(}{\left(}
\renewcommand{\)}{\right)}

\def \lsim{\mathrel{\vcenter
     {\hbox{$<$}\nointerlineskip\hbox{$\sim$}}}}
\def \gsim{\mathrel{\vcenter
     {\hbox{$>$}\nointerlineskip\hbox{$\sim$}}}}
     
\begin{document}

\preprint{DESY 20-177}
\title{Selective enhancement of the QCD axion couplings}

\newcommand{\affBARRY}{{\small \it Physical Sciences, Barry University, 11300 NE 2nd Ave., Miami Shores, FL 33161, USA}}
\newcommand{\affINFN}{{\small \it INFN, Laboratori Nazionali di Frascati, C.P.~13, 100044 Frascati, Italy}} 
\newcommand{\affDESY}{{\small \it Deutsches Elektronen-Synchrotron DESY, Notkestra{\ss}e 85, 
D-22607 Hamburg,  Germany}}

\author{Luc Darm\'e}
\email{luc.darme@lnf.infn.it}
\affiliation{\affINFN}

\author{Luca Di Luzio}
\email{luca.diluzio@desy.de}
\affiliation{\affDESY}

\author{Maurizio Giannotti}
\email{MGiannotti@barry.edu}
\affiliation{\affBARRY}

\author{Enrico Nardi}
\email{enrico.nardi@lnf.infn.it}
\affiliation{\affINFN}

\begin{abstract}
We present a mechanism wherein the QCD axion coupling 
to  nucleons, photons, or  electrons,  can be 
 enhanced selectively  without increasing  the axion 
mass. We focus in particular on the axion-nucleon  couplings, 
that are generally considered to be largely model-independent,
and we show how \emph{nucleophilic} axion models can be constructed. 
We discuss the implications of a nucleophilic axion for astrophysics, 
cosmology and laboratory searches. 
We present a model with enhanced  axion 
couplings to nucleons and photons that can provide an excellent 
fit to the  anomalous emission of hard X-rays recently observed from 
a group of nearby neutron stars, and we argue that such a scenario 
can be thoroughly tested in forthcoming axion-search experiments.

\end{abstract}

\maketitle
	\tableofcontents
	\setcounter{footnote}{0}

\parskip 4pt

\section{Introduction}

With the discovery of Yang Mills instantons \cite{Belavin:1975fg} and of the non-trivial 
vacuum structure of QCD~\cite{Callan:1976je,Jackiw:1976pf} the 
observed absence of CP violation in strong interactions suddenly became  
one of the most serious puzzles of the standard model (SM).
An elegant solution, known as  the Peccei-Quinn (PQ)     mechanism~\cite{Peccei:1977ur,Peccei:1977hh}, was  quickly put forth,
and it is intriguing that, after more than four decades, it is still widely 
considered as the most likely explanation of why CP is a good symmetry of QCD.
A striking consequence of the PQ mechanism is that an ultralight and very feebly coupled pseudo-scalar field, the axion,  must exist~\cite{Weinberg:1977ma,Wilczek:1977pj}.  

In the first and simplest realisation of the PQ mechanism,
the so-called Weinberg-Wilczek (WW) model~\cite{Weinberg:1977ma,Wilczek:1977pj}, 
the  axion couplings to SM fields were not sufficiently suppressed 
and the model  was soon ruled out by laboratory experiments
(for a historical account of  early searches for the WW axion  
see  e.g.~Section~3 in Ref.~\cite{Davier:1986ps}).
Two types of models ensuring a sufficient suppression of all axion couplings 
were then put forth, the Kim-Shifman-Vainshtein-Zakharov 
(KSVZ)~\cite{Kim:1979if,Shifman:1979if} 
and the Dine-Fischler-Srednicki-Zhitnitsky (DFSZ) 
model~\cite{Zhitnitsky:1980tq,Dine:1981rt}. Although these types of axions 
were initially dubbed `invisible' because of the feebleness of their couplings, 
Sikivie showed that search strategies exploiting  the axion coupling to photons
($\gag$) could  still allow to reveal these elusive 
particles~\cite{Sikivie:2020zpn}.  However, three decades of experimental 
efforts have kept probing the axion-photon coupling without yielding 
to its discovery. 
Interestingly,  recent new developments led to the possibility of searching for the axion by 
exploiting  other couplings besides $\gag$. In particular, CASPEr-Wind \cite{JacksonKimball:2017elr} exploits the axion-nucleon coupling 
to search for an axion Dark Matter (DM) wind, originating  from the relative motion  
of the Earth with respect to the Galactic DM halo~\cite{Graham:2013gfa}. 
Another detection strategy is implemented by 
the ARIADNE  collaboration \cite{Arvanitaki:2014dfa,Geraci:2017bmq}, 
which use nuclear magnetic resonance techniques 
to probe an axion-mediated monopole-dipole force, sourced by a macroscopic 
unpolarised material and detected via a polarised sample of nucleon spins. 
Similar approaches involving electron spins are pursued  by 
QUAX-$g_e$ \cite{Barbieri:2016vwg} and QUAX-$g_p g_s$ \cite{Crescini:2017uxs}. 

Presently, the sensitivity of these experiments is 
still far from the parameter space region of  
canonical QCD axion models.\footnote{In the case of ARIADNE, some extra assumptions 
about the structure of CP violation are also required to yield a detectable 
signal.} 
It is then natural to ask whether these experiments  could be already probing  
other types of non-canonical QCD axion models which, although they lie in   
parameter space regions away from the canonical benchmarks, 
can still provide a solution to the strong CP problem.

In this work we discuss a construction wherein the QCD axion coupling to nucleons, photons, or electrons, can be enhanced selectively without increasing 
the axion mass $m_a$.\footnote{By this we mean that the  
QCD relation between $m_a$ and the axion decay constant $f_a$ is not modified. 
This also implies that the axion coupling to the  neutron electric dipole 
moment (nEDM), which depends on this relation, is also unaffected.}  
QCD axion models based on this construction can then populate regions 
in the mass-couplings parameter space which are generally believed 
to be accessible only to (light) axion-like particles (ALPs), 
although a peculiar characteristic of such QCD axions 
is that in most cases  they are endowed with  flavour-violating interactions.
Our construction takes inspiration from the clockwork 
mechanism~\cite{Kim:2004rp,Choi:2014rja,Choi:2015fiu,Kaplan:2015fuy,Giudice:2016yja},
however, it differs from the clockwork axion model 
discussed in Ref.~\cite{Farina:2016tgd} 
in that it introduces  $n+1$ Higgs doublets and a single SM  
singlet complex scalar $\Phi$, and also because, 
similarly to  DFSZ types of scenarios, the QCD anomaly is due 
to the SM quarks rather than to new heavy coloured states.
The construction is in fact more similar to the types of models presented  
in Refs.~\cite{DiLuzio:2017pfr,DiLuzio:2020wdo} in which  
Higgs doublet clockwork gears were used to obtain a  $2^n$ 
enhancement of the axion-photon or axion-electron coupling.

To illustrate the main features of our mechanism we first focus on the axion coupling to  nucleons  ($\gaN$, with $N=p,n$) which are generally considered to be 
 the most model independent of all couplings, and we show that various 
 modifications, and in particular large enhancements, are  instead possible. 
 Note that a first step in the  direction of constructing axion models with modified axion-nucleon couplings  was made in~\cite{DiLuzio:2017ogq}, where it was 
 shown that  variant axion models characterised 
 by generation-dependent PQ charges
 can lead 
 to a strong suppression of $\gaN$. 
 The possibility of  enhancing  $\gaN$ was instead considered in Ref.~\cite{Marques-Tavares:2018cwm}, where  the value of the axion-nucleon coupling was 
 decoupled  from 
that of the axion mass by assigning 
$\U(1)_{\rm PQ}$ charges to SM quarks such that the latter do not contribute to the QCD anomaly. An exponential enhancement of the axion-nucleon coupling is then achieved
via the introduction of several complex scalars $\Phi_k$  hosting in their orbital modes the axion, coupled via a clockwork-like potential.
This construction, however, requires effective dimension five operators in order to 
eventually couple the axion  to the light quarks, while the QCD anomaly of the PQ 
current is  instead due to new KSVZ-like coloured fermions.  
Here we show that a similar result can be obtained with just one SM singlet scalar  
$\Phi$, and without the need of non-renormalizable interactions, 
by introducing additional Higgs doublets. This has also the advantage of 
allowing to enhance different axion couplings rather than $\gaN$.

As regards the nucleophilic axion, the possibility of having 
very light axions with large couplings to the nucleons   implies a rich phenomenology, and opens up 
a parameter space region that can  be largely probed by the 
next generation of axion experiments.
Indeed, 
for axion masses below the $\mu$eV, 
searches at ABRACADABRA 
phase~1 with resonant signal readout~\cite{Kahn:2016aff}, 
and at CASPEr-Wind~\cite{JacksonKimball:2017elr} will cover all scenarios with more than $15$ additional doublets. In particular, ABRACADABRA can have sufficient sensitivity 
to probe models with just $5$ extra doublets for a neV  axion mass. For larger masses, projected sensitivities of KLASH~\cite{Alesini:2019nzq}, 
CAPP~\cite{Semertzidis:2019gkj}, and MADMAX~\cite{Brun:2019lyf} will completely cover the mass range between $\sim 1 \ \mu$eV and $\sim 500 \ \mu$eV. The rest of the parameter space at larger masses could be finally tested at ARIADNE~\cite{Arvanitaki:2014dfa} under the assumption of  maximal CP violation. The next generation axion helioscope IAXO, will also be able to probe a wide mass range~\cite{Armengaud:2019uso} significantly beyond 
the limit from SN1987 cooling, while some regions in parameter might be 
accessible already by BabyIAXO~\cite{Abeln:2020ywv}.  

 Finally, to highlight the flexibility of axion models based on our construction, 
 we address a specific issue which is related to the observation of 
  an excess of hard X-rays emitted from a group of nearby neutron stars (NS)
  referred to  as the ``Magnificent Seven'' (M7)~\cite{Dessert:2019dos}.
  As was argued in Ref.~\cite{Buschmann:2019pfp} interpreting this excess 
   as due to axions produced in the neutron star (NS) core and converted into photons in 
   the NS magnetic field requires a sufficiently light axion mass (below $\sim 10$ $\mu$eV) and at the same time couplings to both nucleon and photons considerably  
   stronger than the ones predicted by canonical QCD-axion models. 
  This is precisely the type of axion that our construction can accommodate, 
  and we show that a nucleophilic axion can in fact provide a very good fit to the observed anomaly.

The paper is organised as follows: in Section~\ref{sec:axionEFT} 
we recall the form of the axion interactions with the SM states
by writing down the relevant effective Lagrangian, and  we introduce the notations.
In Section~\ref{sec:gaN_exp} we describe the details of the construction, 
focusing on the necessary ingredients to obtain a nucleophilic axion,
and illustrate the reasons why  one can generally expect 
flavour violating axion interactions. 
In Section~\ref{sec:experiments}  we explore the phenomenological consequences 
of our scenario and the prospects for  experimental probes in the next-future. 
In Section~\ref{sec:concl} we draw our conclusions.
Two Appendixes complement this paper. 
In Appendix \ref{sec:philic} we present two alternative constructions
yielding respectively a photophilic and electrophilic axion. 
In Appendix~\ref{sec:flavour} we discuss some issues related to 
the structure of the quark Yukawa matrices that can be enforced 
by the  PQ symmetry of our clockwork-inspired multi-Higgs model.

\section{Axion effective Lagrangian} 
\label{sec:axionEFT}

In order to set notations, let us recall the expression of the 
 effective Lagrangian describing the axion interaction 
 with photons and with matter fields 
$f=p,n,e$:
\begin{equation} 
\label{eq:Laint1}
\mathcal{L}_a \supset \frac{\alpha}{8 \pi} \frac{C_{a\gamma}}{f_a} a F_{\mu\nu} \tilde F^{\mu\nu}
+ C_{af} \frac{\partial_\mu a}{2 f_a} \bar f \gamma^\mu \gamma_5 f 
\, ,
\end{equation}
where $\alpha$ is the electromagnetic (EM) coupling constant and
\begin{align}
\label{eq:Cagamma}
C_{a\gamma} &= \frac{E}{N} - 1.92(4) \, , \\
\label{eq:Cap}
C_{ap} &= -0.47(3) + 0.88(3) \, c^0_u - 0.39(2) \, c^0_d - C_{a}
\, , \\
\label{eq:Can}
C_{an} &= -0.02(3) + 0.88(3) \, c^0_d - 0.39(2) \, c^0_u - C_{a}
\, , \\
\label{eq:Cae}
C_{ae} &= c^0_e 
\, , 
\end{align}
with $C_{a} = 0.038(5) \, c^0_s 
+0.012(5) \, c^0_c + 0.009(2) \, c^0_b + 0.0035(4) \, c^0_t$
the sea quarks contribution.  
In \eqn{eq:Cagamma} 
$E$ and $N$ 
are respectively the EM and QCD anomaly coefficients that are defined in terms of the anomalous PQ current 
\beq 
\label{eq:dJPQAnomal}
\partial^\mu J^{\rm PQ}_{\mu} = 
\frac{\alpha_s N}{4\pi} 
G^a_{\mu\nu} \tilde G^{a{\mu\nu}}
+\frac{\alpha E}{4\pi} 
F_{\mu\nu} \tilde F^{\mu\nu}
\, , 
\eeq
with $\alpha_s$ the strong interaction coupling constant. While the axion couplings to the quarks $c^0_q$ with  $q = u, d, s, c,b,t$ 
appearing in \eqn{eq:Cap} and (\ref{eq:Can}) 
are defined by the Lagrangian term
\beq 
\label{eq:fermionc}
c^0_q \frac{\partial_\mu a}{2 f_a} \bar q \gamma_\mu \gamma_5 q \, .
\eeq 
Taking a Yukawa term $\bar q_L q_R H_q$, 
a simple expression for $c^0_q$ in terms of PQ charges is \cite{DiLuzio:2020wdo} 
\beq
\label{eq:c0fPQcharges}
c^0_q = \frac{\mX_{q_L} - \mX_{q_R}}{2N}  
= \frac{\mX_{H_q}}{2N} \, ,   
\eeq
where in the last step 
we have replaced the fermion PQ charges $\mX_{q_{L,R}}$ with the charge $\mX_{H_q}$ of the corresponding Higgs doublet. 
Finally, a common way to rewrite the axion interactions arising from \Eqn{eq:Laint1} which will be used  in Section~\ref{sec:experiments} is
\beq 
\label{eq:Laint2}
\mathcal{L}_a \supset 
 \frac{1}{4} g_{a\gamma} a F_{\mu\nu} \tilde F^{\mu\nu}
- i g_{af} a \bar f \gamma_5 f \, ,
\eeq
where 
we have defined 
\beq 
\label{eq:gagammagaf}
g_{a\gamma} = \frac{\alpha}{2 \pi} \frac{C_{a\gamma}}{f_a} \, , \qquad 
g_{af} = C_{af} \frac{m_f}{f_a} 
\, . 
\eeq

\section{Enhancement mechanism and nucleophilic axions} 
\label{sec:gaN_exp}

Enhancing selectively 
a given coupling of the axion typically requires a mechanism to sizeably increase the axion coupling to the nucleons, to the electrons or to the photons, 
without increasing at the same time the coefficient of the QCD 
anomaly. In this section, we will illustrate how this can be 
realised in order to generate a nucleophilic axion.

\subsection{Generation-dependent PQ charges}
\label{sec:gaN_exp_idea}

The key ingredient of our construction is the existence of a large 
hierarchy among the PQ charge differences $\mX_{q_L} -\mX_{q_R}$  
for the quarks of different generations. 
We will start by assuming that the charge 
differences for the first generation have hierarchically large values. 
The overall contribution of the first generation to the 
coefficient of the QCD anomaly, however,  vanishes if  
the value for the up quark is equal in size 
but opposite in sign to that of the down quark. 
The  anomaly coefficient then remains determined by the 
charges of the quarks of the other two generations, which we assume 
to have $O(1)$ charge differences. Given that for each quark flavour 
the L-R charge differences must match the charge of the Higgs doublet 
responsible for the mass of that specific quark, see \eqn{eq:c0fPQcharges}, 
it is clear that to realise this scenario the Higgs sector must be extended 
to include additional Higgs multiplets. 
Hence we start by assuming that the fermion Yukawa couplings 
involve three Higgs doublets $H_0,H_1$ and $H_n$ with  hypercharge $Y=-\frac{1}{2}$ and PQ charges $\mX_0, \mX_1$ and $\mX_n$. For the third Higgs doublet we assume a hierarchically large charge value:
\begin{align}
\label{eq:Xn}
   \mX_n \ \gg \  \dchi \equiv \mX_1 - \mX_0 \ ,
\end{align}
where, without loss of generality,  
we have taken the difference  between the  first two PQ charges to be positive, 
$ \mX_1 - \mX_0 >0 $. 
A detailed mechanism  that can produce the charge  hierarchy   in \eqn{eq:Xn}
will be discussed in Section~\ref{sec:clockwork}.
Next, we  assume that the three Higgs doublets  couple to the quarks  
via the following generation-dependent  Yukawa operators:
\begin{align}
\label{eq:gaNexpHL}
\nonumber
&\bar u_{L} u_{R}  H_n + \bar d_{L} d_{R}  \tilde H_n  +
\bar c_{L} c_{R}  H_0 + \bar s_{L} s_{R}  \tilde H_0  + \\
&\bar t_{L} t_{R}  H_1 + \bar b_{L} b_{R}  \tilde H_0  \,, 
\end{align}
where $\tilde H = i \sigma_2 H^*$.
This structure realises the conditions described above: 
the axion couplings to the $u$ and $d$ quarks are proportional 
to the \emph{large} charge $\mX_n$
\begin{equation}
\label{eq:expPQupL}
c^0_{d} = - c^0_{u} = \frac{\mX_{d_L}-\mX_{d_R}}{2 N} = -\frac{\mX_{n}}{2N} \,,  
\end{equation}
and they do not get particularly suppressed by the coefficient of 
the QCD anomaly that it is fixed in terms of the \emph{small} charges 
of the quarks of the third generation:
\begin{align}
\label{eq:expgaN_anomalyL}
2N &=\(\mX_{t_{L}} - \mX_{t_{R}} \)
+ \(\mX_{b_{L}}-  \mX_{b_{R}} \)  = \dchi \,. 
\end{align}
Consequently, also the axion couplings to the nucleons in \eqns{eq:Cap}{eq:Can} 
get significantly enhanced by the large couplings of the light quarks
$|c^0_{d}| = |c^0_{u}| = \mX_n/\dchi \gg 1$.

The Cabibbo-Kobayashi-Maskawa (CKM) mixing angles can be generated 
by adding to \eqn{eq:gaNexpHL} inter-generational operators 
as for example $(\bar c_{L} u_{R}  H_1)+
(\bar t_{L} u_{R}  H_n)$, together with additional terms 
consistent with the charge
assignments implied by the latter two operators plus those in Eq.~\eqref{eq:gaNexpHL}. 
We anticipate that the axion will be eventually contained 
in the neutral `orbital modes' of the doublets $H_{0,1,n}$ and thus, since 
the charge assignments in Eq.~\eqref{eq:gaNexpHL} are generation-dependent, 
once the quarks are rotated to their mass basis
the axion couplings to the quarks will in general be flavour-violating,
see Section~\ref{sec:fcnc}.

As regards the leptons, they couple to the complex 
conjugate Higgs doublets  $\tilde H_{0,1,n}$.  
The EM anomaly coefficient $E$ is then readily obtained as:
\begin{align}
\label{eq:EoNgeneralDFSZ}
E &= \sum_j \left( 
 \frac{4}{3}  \mX_{H_{u_j}} -
  \frac{1}{3}  \mX_{H_{d_j}} -
     \mX_{H_{e_j}}\right) \nonumber \\
 &=\frac{8}{3} N + \sum_j \left( \mX_{H_{d_j}}- \mX_{H_{e_j}} \right) \, .
\end{align}
It is clear that depending on which specific doublet 
the leptons are coupled to, the 
EM anomaly could be also enhanced by the large charge $\mX_n$, 
or could  remain of the order of the QCD anomaly.
For the nucleophilic axion we are interested in the latter possibility,   
so we will assume that the leptons couple universally to $\tilde H_0$: 
\beq 
\label{eq:LYlep}
\bar \ell_{iL} e_{iR} \tilde H_0\qquad (i=1,2,3) \, . 
\eeq
The  $E/N$ factor is then given by:
\begin{align}
\label{eq:EoN1}
\frac{E}{N}  &=\frac{8}{3} + 2\frac{\X_n - \X_0}{\dchi} \, .
\end{align}

It is important to remark at this point that  the pattern 
of Yukawa couplings in \eqn{eq:gaNexpHL}  can be straightforwardly 
re-arranged in a different way to yield:
\begin{itemize}
    \item An axion dominantly nucleophilic: $\gaN \gg \gae, \gag$. This can be obtained for instance by coupling the up and down quarks and the $\tau$ lepton to $H_n$ and $ \tilde{H}_n$
    respectively.  We obtain $E , N\sim \mathcal{O}(1)$ but enhanced coupling to nucleons and to the $\tau$ lepton.
    \item An axion dominantly photophilic: $\gag \gg \gaN, \gae$. This is easily obtained by coupling for instance \textit{only} the third generation leptons to $\tilde{H}_n$. We obtain $E \propto \mX_n$, $N\sim \mathcal{O}(1)$ and enhanced coupling to the~$\tau$.
    \item An axion dominantly electrophilic: $\gae \gg \gaN, \gag$. This can be obtained by coupling among the leptons \textit{only} the electron to $\tilde H_n$, and 
    for example $b$ and $c$ respectively to $\tilde H_n$ and  $H_n$. In this way  the `large' contributions to $E$ and $N$ cancel out, and  $\gan$ is not enhanced.  
    We obtain $E , N\sim \mathcal{O}(1)$ but enhanced coupling to $e$ (and to $b$ and $c$).
\end{itemize}
Two examples of nucleophilic axions will be discussed in Section~\ref{sec:couplings},
while two model realisations for a photophilic and an electrophilic axion 
will be presented in Appendix~\ref{sec:philic}. 

The large hierarchies in the couplings described above imply that 
large radiative contributions to  suppressed couplings are possible. 
For example, a large $\gag$ would generate, via a triangular loop, a large radiative 
contribution to the lepton couplings that could be relevant when 
the tree-level value $c^0_\ell$ is not particularly large~\cite{Srednicki:1985xd,Chang:1993gm}
\begin{align}
\label{eq:c0e}
    \delta c_{\ell}^0 &= \frac{3 \alpha^2 Q^2_\ell}{4 \pi^2} \left[ \frac{E}{N} \log \left( \frac{f_a}{\mu_{\rm IR}}\right) + \dots   \right] \nonumber \\
    & \sim \frac{3 \alpha Q^2_\ell}{2 \pi} \gag \log \left( \frac{f_a}{\mu_{\rm IR}}\right) + \dots   \ ,
\end{align}
where $Q^2_\ell=1$,  $\mu_{\rm IR}$ is the IR scale 
at which the coupling is evaluated and the second equality holds for $E/N \gg 1$. 
Radiative corrections to $\gae$ from large axion couplings to the quarks  
are instead a two-loop effect and are more suppressed. 
Similarly to the leptons, the axion coupling to the quarks would also receive 
a large radiative contribution from an enhanced $\gag$
analogous to the one in \eqn{eq:c0e}. 
The converse is instead not true: 
large axion-couplings to the fermions 
do not yield large radiative contributions to $\gag$. 
This is because in the effective field theory limit $m_f \to \infty$, the 
axion-photon coupling is solely fixed in terms of the ratio of 
anomaly coefficients $E/N$ and does not renormalize. 
The contribution of  fermion loops in fact 
requires a helicity-flip, and for finite $m_f$ is suppressed 
as $m^2_a / m^2_f$ and thus completely negligible.
Finally,  when the couplings of the first generation quarks  $c^0_{u,d}$
are not particularly large, while at least one quark of the heavier generations 
has a much larger coupling, the axion-nucleon coupling $\gaN$ 
might become dominated by the sea quark contribution $C_a$. 
From the expression of $C_a$ given below \eqn{eq:Cae}
it can be seen that if $c^0_t/c^0_{u,d} \gsim 250$ even the contribution 
of the top quark would exceed that of the  valence quarks.  \\

\subsection{Flavour violating axion couplings}
\label{sec:fcnc}

The set of Yukawa operators in \eqn{eq:gaNexpHL}
implies that the corresponding model belongs to the class of  
multi-Higgs doublet models with  no natural flavour 
conservation~\cite{Glashow:1976nt}, in which 
the exchange of Higgs scalars can represent  a dangerous 
source of flavour changing neutral currents (FCNC). 
These effects, however,  can be safely suppressed 
by assuming the so-called \emph{decoupling limit}~\cite{Haber:1989xc}  
which ensures that a single neutral Higgs scalar  with 
properties indistinguishable from that of  the SM-Higgs boson
survives in the low-energy spectrum, so that  
 the model can be rendered consistent with 
 limits on FCNC processes, with the LHC measurements  of Higgs properties as well as with electroweak precision measurements.
The decoupling limit generically requires 
that a set of dimensional parameters in the 
scalar sector has values much larger than 
the electroweak scale,  and yet, to ensure 
a light (electroweak-scale) neutral Higgs, the  
determinant of the  matrix of the neutral scalar squared 
masses should vanish in the limit $v^2 \to 0$. 
Clearly,  this requires a tree level fine-tuning 
among the parameters, besides the usual one between the tree-level Higgs mass and the quadratically divergent loop contributions.

Another source of flavour violation which does not decouple in 
the same limit is represented by flavour violating axion interactions,
which appear when the quark fields in \eqn{eq:gaNexpHL} are rotated to 
the mass basis. Besides axial-vector couplings 
analogous to the ones given in~\eqn{eq:fermionc}, 
off-diagonal  axion-quark interactions are  characterised 
also by vector interactions. 
In matrix notation, the axion-quark couplings can be written as
\begin{equation}
    \label{eq:offdiagcurrents} 
-\frac{\partial_\mu a}{2 f_a} \left[\bar q \gamma^\mu \left(c_q^{0V} 
- c_q^{0A} \gamma_5\right) q\right]\,, 
\end{equation}
where  $q$ are vectors containing the 
three types of up- or down-type quarks, and 
$c^{0V,A}_{q}$ (with elements 
$c^{0V,A}_{q_i q_j}$) are  $3\times 3$ 
matrices  of couplings  defined as:
\begin{align}
\label{eq:c0qfv}
 c^{0 V,A}_q  =  \frac{1}{2 N} \left( {U^q_L}^\dagger 
 \mXb_{q_L} U^q_L \pm {U^q_R}^\dagger \mXb_{q_R} U^q_R \right) \ ,
\end{align}
where $\mXb_{q_{L,R}}$ are diagonal matrices of the 
PQ charges of the quarks, e.g.~$\mXb_{u_L} =(\mX_{u_L}, \mX_{c_L},\mX_{t_L})^T$, and $U^{q}_{L,R}$ are the quark unitary rotation matrices.
Note that 
flavour violating effects will depend on 
the differences between the charges of the same-type (up- or down-,  L or R) 
quarks of different generations, and get enhanced when these differences are large. 
 The most relevant bounds on axion-quark flavour-violating effects 
 arise from mesons decays,  which yield the following limits~\cite{Bjorkeroth:2018ipq,MartinCamalich:2020dfe}:
\begin{align}
\label{eq:limflav}
f_a \gtrsim \{ 3.4 \cdot 10^{11} c^V_{sd},  \ 
1.7 \cdot 10^{8} c^V_{bs}  , \  
6 \cdot 10^{7} c^V_{bd} \}\,\mathrm{GeV} \,,
\end{align}
 where for simplicity we have used  for the quark mass eigenstates 
 the same labels $q=d,s,b,\dots$.
With the charge assignment implied by Eq.~(\ref{eq:gaNexpHL}),  
we can use Eq.~\eqref{eq:c0qfv}, obtaining for example
the down-type quarks of  
the first and second generation: 
\begin{align}
c^V_{sd} = \frac{\mX_n-\mX_{0}}{2N}  \left( U^{d_L \dagger}_{11} 
U^{d_L}_{12} x_{d_L} + U^{d_R \dagger}_{11} U^{d_R}_{12} x_{d_R} \right) \,,
\end{align}
where we have defined $x_{d_L} = (\mX_{d_L}-\mX_{0})/(\mX_{n}-\mX_{0})$ and  similarly for $x_{d_R}$. 
A suppression of $c^V_{sd}$ can be straightforwardly obtained 
for example by  fixing $U_{12}^{d_L} \sim U_{12}^{d_R} \sim 0$ and 
generating the Cabibbo angle 
solely from the up-quark sector, 
that is by assuming that the large PQ charges are associated 
with particularly small mixing angles. 
In the following, we work for simplicity  in the approximation in which 
inter-generational mixing effects can be neglected,  
so that the current basis coincides, to a good approximation,  
with the mass basis.\footnote{Quark mixing also induces  
flavour-diagonal corrections to the axion couplings~\cite{DiLuzio:2017ogq}.
The coupling to a heavy quark $q_i$ (with PQ charge $\mX_{0}$ or $\mX_{1}$) will 
for example receive a correction $\delta c_{q_i}$ proportional to the light quark charges, 
which can in principle concur to reduce the hierarchies between the various quark 
couplings. We will neglect this effect in this work.}

\subsection{Clockwork enhancement of the PQ charges}
\label{sec:clockwork}

In order to generate a hierarchy in the PQ charges 
 as given in \eqn{eq:Xn}
we add to the three doublets $H_{0,1,n}$ 
an additional set of $n-2$ scalars $H_2, H_3,\dots, H_{n-1}$, for a total of  $n+1$ Higgs doublets all with hypercharge 
$Y=-\frac{1}{2}$.\footnote{no Landau poles below the Planck scale, and assuming conservatively a unique threshold of the order of the electroweak scale for the contribution of all the new scalars, still results in a fairly large 
limit on the allowed number of doublets $n \lesssim 50$.} 
Besides the extra  doublets, we also introduce 
an electroweak singlet scalar field $\Phi$ with 
PQ charge $\X_\Phi$ and vacuum expectation value (VEV) $v_\Phi \gg v$. 
We assume that $\Phi$ couples to $H_0$ and $H_1$ via one of the following 
two renormalizable  terms 
\beq
\label{eq:HPhi}
H_1^\dagger H_0 \Phi  \qquad \textrm{or} \qquad H_1^\dagger H_0 \Phi^2 \,, 
\eeq
so that $\dchi =  \mX_\Phi$ or  $\dchi = 2 \mX_\Phi$. 
With the first choice $2N=\mX_\Phi$   (see \eqn{eq:expgaN_anomalyL}),  
 the QCD potential has the same periodicity than the axion field, 
hence there is a single potential  minimum and the number 
of domain walls (DW)~\cite{Sikivie:1982qv} is $\NDW=1$. With the second 
choice $2N=2\mX_\Phi$,  there are two physically distinct but degenerate minima, 
and $\NDW=2$. 
It should be noted that this result crucially depends on the fact 
that the quarks that determine the anomaly and the field $\Phi$ 
couple to the same pair of Higgs doublets $H_{0,1}$, and it would not hold 
if, for example, $\Phi$ is coupled to a different pair of  doublets 
or if,  maintaining the scalar couplings in \eqn{eq:HPhi}, 
$\mX_n$ contributes to the QCD anomaly.  

The $n+2$ scalar fields $\{H_i,\Phi\}$ carry a $U(1)^{n+2}$ 
rephasing symmetry $H_k \to e^{i \alpha_k} H_k$,  
$\Phi \to e^{i \alpha_\Phi} \Phi$. We will assume that, in addition to 
one of the operators in \eqn{eq:HPhi},   the scalar potential also contains 
the following set of quadrilinear terms
\begin{equation}
\label{eq:gaNclock}
(H^\dagger_{k-1} H_k)(H^\dagger_{k-1} H_0) \,, \quad  k=2,\dots,n\,,
\end{equation}
so that  $U(1)^{n+2}$ is broken explicitly to $U(1)_{\rm PQ}\times U(1)_Y$.
Since $H_{0,1,n}$ need to pick-up a VEV to generate the quark masses, 
even in the case when all the additional doublets 
$H_{k}$  ($k=2,\dots,n-1$) have positive mass square terms, they will still 
acquire induced  VEVs  because  they appear linearly in the terms in \eqn{eq:gaNclock}.
This feature can be used to generate a significant hierarchy between the VEVs. 
Indeed, assuming that   $H_{0,1}$  acquire respectively  the VEVs  $v_{0}$ 
and $v_1$, and 
that all the other doublets have positive mass square terms $\mu^2_k >0 $ 
the   induced VEVs will read 
\beq
\label{eq:vevrec}
v_k \sim \frac{v_{k-1}^2}{\mu_k^2}\,v_0\,,  \quad k\geq 2\,, 
\eeq
so that small VEVs can be typically expected if the masses of the $k>2$ doublets are large
$\mu_k \gg v_{0,1}$ or if the couplings of the operators in \eqn{eq:gaNclock}
are small. 

The arrangement of the quadrilinear Higgs couplings in \eqn{eq:gaNclock} 
implies that  the PQ charges  $\X(H_k) = \X_k$ satisfy:
\beq
\label{eq:Xiter}
\mX_k = 2^{k-1} \dchi + \mX_0\,, \qquad k=2,\dots,n  \, , 
\eeq
that is $\mX_n $ is exponentially enhanced with respect to   
$\dchi $.\footnote{
A nearer neighbour set of operators 
$(H^\dagger_{k-1} H_{k})(H^\dagger_{k-1} H_{k+1})$ 
would also imply exponentially enhanced PQ charges 
$\mX_k = \frac{1}{3}[1- (-2)^k] \dchi + \mX_0$. However, 
 \eqn{eq:Xiter}  has a simpler form and hence 
throughout this paper we will assume the scalar couplings 
in \eqn{eq:gaNclock}.}

In the presence of many Higgs doublets carrying PQ charges, identifying the  
{\it physical} axion and deriving its couplings to the fermions involves some subtleties. 
To ensure that the axion has no component in the longitudinal mode of the $Z$ boson,  
one has to impose  an  orthogonality condition between the 
PQ and hypercharge currents  
  $J^{\rm PQ}_{\mu}|_a   =\sum_i \mX_i v_i \partial_\mu a_i $ 
  and    $J^{\rm Y}_{\mu} |_a = \sum_i Y_i v_i \partial_\mu a_i $,  
  where the sum runs over all the scalar doublets ($i=0,1,\dots,n $),  
$v_i=\sqrt{2}\vev{H_i}$
  are their VEVs,  and   $a_i$ are the neutral orbital modes of the Higgs fields. 
The orthogonality condition reads 
\beq
\label{eq:orthogonalityHk}
0 = \sum_{i=0}^{n} 2 Y_i  \mX_i v_i^2 = \mX_{0} v^2 
+ \dchi
\sum_{j=1}^{n} 2^{j-1} v_j^2 \, ,
\eeq
where $v^2 = \sum_{i=0}^n v^2_i 
\approx 246$ GeV is the electroweak breaking VEV. 
From  \eqn{eq:orthogonalityHk} we see that  
the values of the PQ charges are determined in terms 
of the  charge difference $\dchi$ and of the structure of 
the Higgs doublets VEVs.
Let us define
\begin{equation}
\label{eq:kappa}
\kappa  \equiv  \sum_{j=1}^n \ \frac{2^{j}}{2^n} \; \frac{v^2_j}{v^2}  \,.
\end{equation}
$\mX_0$ and $\mX_n$ can then be written as
\begin{eqnarray}
\label{eq:X0clockwork}
\mX_{0} &=& 
 -\dchi \sum_{j=1}^{n} 2^{j-1}\; \frac{v_j^2}{v^2} = 
- 2^{n-1}\, \dchi\, \kappa
\, , \\
\label{eq:Xnclockwork}
\mX_n  &=& 
2^{n-1}\, \dchi\, (1-\kappa)\, ,
\end{eqnarray}
where the second equation makes use of \eqn{eq:Xiter}. 
 From these  two equations we obtain 
  \begin{equation}
     \label{eq:XnXo}
     \frac{\mX_n}{\mX_0} = \frac{\kappa-1}{\kappa}
      \end{equation}
which makes clear that  the hierarchy between the PQ charges 
is determined by the parameter $\kappa$. 
In principle $\kappa$ could range in the interval $[0,1]$ with the smallest 
values  requiring a strong suppression of the VEVs of the doublets 
with the largest PQ charges. However, as we will see in the next 
Section, the  phenomenologically allowed range is in fact slightly narrower. 
The value of $\kappa$ will be crucial 
to determine the values of the axion couplings. Note 
that the anomaly coefficients $E,\,N$ do not depend on $\kappa$,  
as can be explicitly verified  by inserting 
in  \eqn{eq:EoN1} $\mX_n-\mX_0 = 2^{n-1}\dchi$.
Namely, $E,\,N$  are insensitive to the particular VEV structure, as could 
have been expected  since  anomalies do not depend on IR physics.

The physical axion field is defined 
in term of the VEVs of the Higgs doublets and of the electroweak singlet 
and of their neutral pseudo-scalar components 
$a_i$ and $a_\Phi$  as 
\begin{align}
    a & = \frac{1}{v_a}
    \left( \mX_\Phi v_\Phi a_\Phi + \sum_{i=0}^n \mX_i v_i a_i\right)   \\ 
    v_a^2 & = \mX_\Phi^2 v_\Phi^2 + \sum_{i=0}^n \mX_i^2 v_i^2 \,. 
\end{align}
Note that due to the exponential enhancement of the Higgs  charges $\mX_i$ 
for large values of $i$, it might not be always accurate to approximate  
$a\approx a_\Phi$.  
What remains true, is that the scale that suppresses all axion couplings 
is bounded from below by $v_a > v_\Phi$, and hence for sufficiently large 
values of $v_\Phi$ all current limits on the axion couplings can be 
easily evaded.

\subsection{Axion couplings to matter}
\label{sec:couplings}

The axion coupling to the SM quarks and leptons  depend on the particular 
Higgs doublet to which the  fermion couples, 
and on the value of $\kappa$, that  is on  the vacuum structure.
From \eqn{eq:gaNexpHL} and \eqn{eq:LYlep}
we see that the second generation quarks, the $b$-quark and the 
leptons interact with $H_0$. Their couplings to the axion 
are then given by (see \eqn{eq:X0clockwork}):
\beq 
\label{eq:ccsb}
c^0_{c} = -c^0_{s,b,\ell} =  \frac{\mX_{0}}{2 N}  
= - 2^{n-1} \kappa \, 
\eeq
The  quarks of the first generation couple to $H_n$, so that from 
\eqn{eq:Xnclockwork}  we have  
\beq 
\label{eq:cud}
 c^0_{u} = -c^0_{d} =
 \frac{\mX_{n}}{2 N}  
= 2^{n-1} (1-\kappa) \,.
\eeq
We see from these equations that the parameter $\kappa$ is crucial to determine a hierarchy in the couplings strength, especially if its value can approach 
the boundaries of the interval $[0,1]$. 
Let us denote by $v_f$ the VEV of the Higgs doublet coupled to the fermion $f$.
Perturbativity of the Yukawa couplings $y_f \lesssim \sqrt{4\pi}$  requires 
\beq
\label{eq:pertvev}
 v_f \gtrsim \frac{m_f}{\sqrt{2\pi}} \, .
\eeq
Therefore, while $\kappa$ could in principle vanish for $v_0 \to v$, such configuration is clearly forbidden by non-zero quark masses.  In particular, from \eqn{eq:gaNexpHL} and (\ref{eq:pertvev}) the perturbativity of the top Yukawa in the $n+1$ Higgs theory 
translates into $(v_1/v)^2 \gtrsim (m_t / (v \sqrt{2\pi}))^2 \approx 0.08$. 
Then a lower bound on $\kappa$ can be readily obtained by retaining  only 
the first term of the sum \eqn{eq:kappa}
\beq 
\kappa \; \geq\; \frac{2}{2^n} \frac{v_1^2}{v^2} \; \gtrsim \; \frac{0.08}{2^{n-1}}\,.
\eeq
Due to the suppressing exponential factor $2^{n-1}$ the lower value for  
$\kappa$ does not depart significantly from zero. 
Note that the only contribution to $\kappa$ that has no exponential suppression 
comes from $v^2_n/v^2$, but for this contribution the perturbative limit 
comes from the down quark mass, and it remains below $10^{-10}$. 
As was remarked after \eqn{eq:XnXo}, the maximum value of $\kappa$ 
is obtained when $v_n$ is maximum, that is when $v^2_n \approx v^2-v^2_1$.
From this we obtain the upper limit
\beq 
\kappa \; \lesssim \; \frac{2}{2^n} \frac{v_1^2}{v^2} + \left(1-\frac{v_1^2}{v^2}\right) \;
\lesssim \; 0.92 
\eeq
where the last relation holds in the limit of large $n$. All in all the 
phenomenologically allowed range for $\kappa$ is 
\beq
\label{eq:krange}
0< \kappa \lsim 0.92\,.
\eeq
In terms of $\kappa$ the axion couplings to the SM fermions and to the photon read (model A): 
\begin{align}
\label{eq:gagA}
g_{a\gamma}&\simeq  2^n  \,  \alpha/({2\pi f_a})\,, \\ 
    \label{eq:gapA}
g_{a p} &\simeq   2^{n-1}\times 1.27\,   (1-1.02\kappa) \,  (m_p/f_a)\,, \\
\label{eq:ganA}
\gan &\simeq -  2^{n-1}\times  1.27\, (1-0.98\kappa) \,  (m_n/f_a) \,, \\
\label{eq:gaeA}
g_{ae} &= \left(2^{n-1}\, \kappa +\delta c_\ell^0\right)  \, (m_e/f_a)\,.
\end{align}
In order to highlight the exponentially enhanced contributions, in writing 
these equations we have made some approximations: 
in the first three relations we have neglected the  
model independent contributions to the couplings (the pure numbers in the left-hand side of  \eqs{eq:Cagamma}{eq:Can}) which are clearly subdominant, and  
for the axion-photon coupling we have also omitted the 
 factor of $8/3$  appearing in \eqn{eq:EoNgeneralDFSZ}. However, 
in Eqs.~\eqref{eq:gapA} and \eqref{eq:ganA}  we have included the $s$ and $c$  sea quark contributions which can also get exponentially enhanced.
Finally, let us recall that the couplings to the fermions  also receive a 
radiative contribution from triangle loops involving $\gag$.
In view of the upper limit on $\kappa$ in \eqn{eq:krange} this correction is irrelevant for the axion-nucleon couplings so it has been neglected in the expressions for $\gap$ and  $\gan$,
but it can become important for $\gae$ in the limit $\kappa \to 0$, so  in \eqn{eq:gaeA} 
we have included the corresponding correction $\delta c_\ell^0$ which is 
given in \eqn{eq:c0e}.

For generic values of $\kappa$ within the range given in \eqn{eq:krange} 
the  structure of the couplings in \eqs{eq:gagA}{eq:gaeA} 
naturally favours an enhancement of  the axion interaction with the nucleons
and with the photon. 
However, as was anticipated in Section~\ref{sec:gaN_exp_idea}, 
we can easily arrange a  pattern of Higgs-fermion  couplings different 
from the ones given in \eqn{eq:gaNexpHL} and \eqn{eq:LYlep},   
and produce  other types of unconventional axions,  for instance dominantly  photophilic or dominantly electrophilic. Since the corresponding models can 
also be of phenomenological interest, we discuss some examples 
in  Appendix~\ref{sec:philic}.
Here we discuss the possibility of generating 
a certain suppression of the axion couplings to the nucleons  
with  respect to the coupling to the photon, which in 
\eqs{eq:gagA}{eq:ganA} have similar enhancements. 
This might in fact be desirable in view of the strong limit 
on $\gaN$ from the duration of the SN1987A neutrino
burst~\cite{Carenza:2019pxu} and, in particular, it is required 
if one attempts to fit the observed excess 
of hard X-ray emission from a group of nearby NS~\cite{Buschmann:2019pfp} in terms of axion emission from the NS core,   
a possibility that  will be analyzed quantitatively in Section~\ref{sec:experiments}.
A simple way to suppress to a certain extent $\gaN$ is to couple the `large charge' Higgs  $H_n$ to second generation quarks.
The axion-nucleon interaction then receives the dominant 
contribution from the $s$ and $c$ sea quarks, which have additional  
suppression factors, and  we have (model B):
\begin{align}
\label{eq:gagB}
g_{a\gamma}&\simeq  2^n  \,  \alpha/({2\pi f_a})\,, \\ 
    \label{eq:gapB}
g_{a p} &\simeq   2^{n-1}\times 0.026\,   (1-50\kappa) \,  (m_p/f_a)\,, \\
\label{eq:ganB}
\gan &\simeq  2^{n-1}\times  0.026\, (1+48\kappa) \,  (m_n/f_a) \,, \\
\label{eq:gaeB}
g_{ae} &= \left(2^{n-1}\, \kappa+\delta c_\ell^0\right)  \, (m_e/f_a)\,, 
\end{align}
with  in this case $10^{-6}\lesssim \kappa \lesssim 0.92$,  
where the lower limit corresponds to require  that the 
charm Yukawa coupling remains perturbative.\footnote{We use 
conservatively the values of the quark masses at the electroweak scale  
computed in the $\bar{\rm MS}$ scheme.}

\section{Phenomenological implications} 
\label{sec:experiments}

The models discussed above have a rich phenomenology. 
The enhanced coupling to nucleons allows experimental searches through novel methods
that have been  put forth in recent years, and it also implies an efficient productions 
in stellar environments at low values of the axion mass. 
Both models A and B in fact predict 
 enhanced couplings to the photons and, depending on the choice of the 
 parameter $\kappa$, the couplings to electrons might get enhanced as well,
 so that these models can be called \emph{astrophilic}.
To satisfy the good agreement between stellar evolutionary models 
and observations, the axion decay constant $f_a$ should then be
increased to very large values to counterbalance 
the effects of the large couplings induced by the exponentially 
enhanced PQ charges. 
Thus, stellar evolution forces astrophilic  axions to be unusually light.
For example, astrophysics bounds allow  the well-studied DFSZ axion
to have a mass up to about 10~meV, 
whereas for our nucleophilic axions, for $n\gtrsim 15$ 
the mass is constrained to lie below 1~$\mu$eV  (see Figs.~\ref{fig:astro_bounds_A} and \ref{fig:astro_bounds_B}).
The large couplings/small mass feature of these models is quite interesting from 
the experimental prospective since, as will be discussed in 
Section~\ref{sec:axionexp}, several proposed experiments can  have 
enough sensitivity to probe large portions of  the low mass region 
not yet excluded by astrophysics observations.

\subsection{Astrophysics} 
\label{sec:astro}

\begin{figure}[t]
\centering
\includegraphics[width=8.7cm]{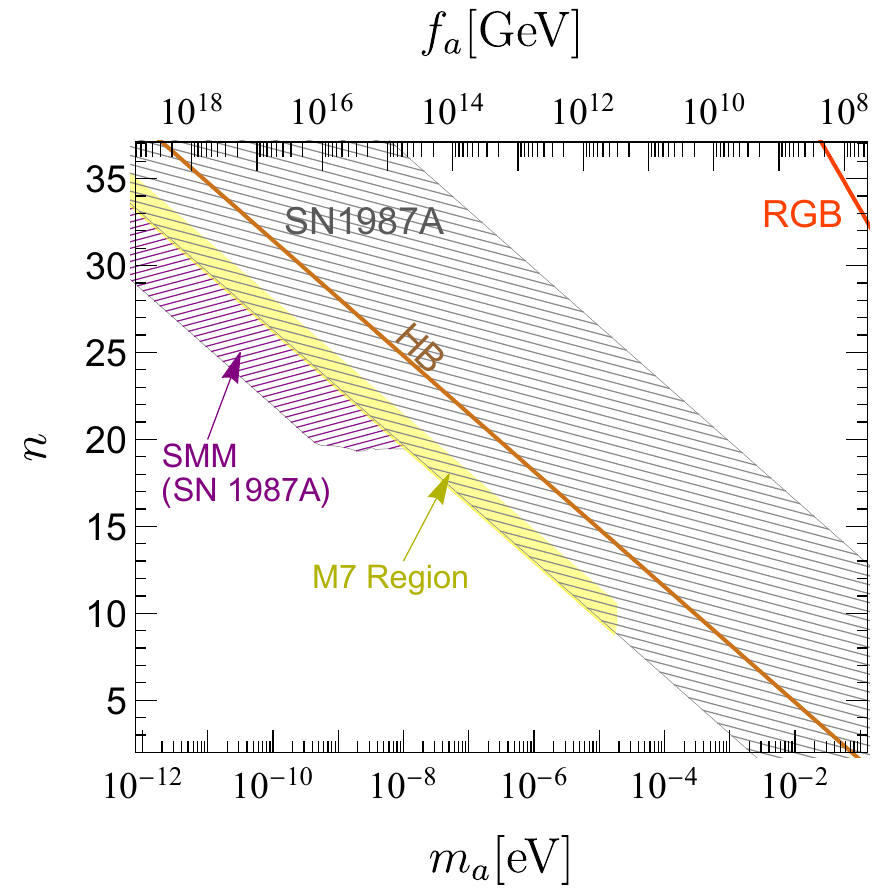}
\caption{Astrophysical bounds on model A for $\kappa \sim 0$
(strong suppression of $\gae$).  
The region hatched in grey is excluded by the SN1987A neutrino burst duration, 
the purple hatched region is excluded by the non-observation of 
gamma rays from conversion of SN1987A axions. 
The HB and  RGB excluded regions lie above the corresponding lines. 
The  $2\sigma\,$c.l.~region from a fit to the M7 anomaly  is 
depicted in yellow and is completely excluded by the SN1987 bound.
}
\label{fig:astro_bounds_A}
\end{figure}

\begin{figure}[t]
\centering
\includegraphics[width=8.7cm]{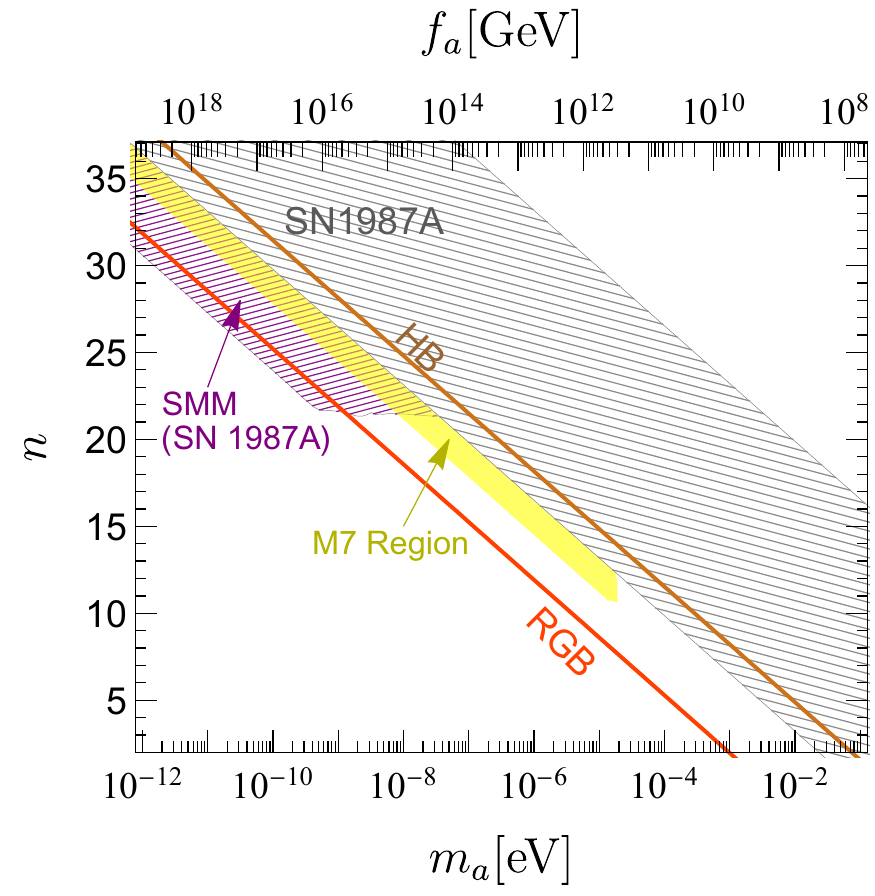}
\caption{Same as Figure~\ref{fig:astro_bounds_A}
but for $\kappa = 0.92$ ($\gae$ exponentially enhanced).
Part of the M7 yellow region is compatible with the SN1987 bound, 
but it is now firmly excluded by the RGB limit.
\label{fig:astro_bounds_B}
}
\end{figure}

 Stellar evolutionary theoretical studies, combined with accurate observations of stellar populations, lead to strong constraints on the axion  couplings to matter 
 and radiation (see, e.g., \cite{Raffelt:1996wa,Giannotti:2015kwo,Giannotti:2017hny,DiLuzio:2020wdo,DiLuzio:2020jjp}).
The bounds from stellar evolution  for model A, 
corresponding to the couplings 
in \eqs{eq:gagA}{eq:gaeA}, are shown in Fig.~\ref{fig:astro_bounds_A} 
and \ref{fig:astro_bounds_B}.  
In these plots the $x$-axis corresponds to the 
axion mass, and the $y$-axis to the number of Higgs doublets 
(i.e.~the number of clockwork gears)  that in our construction are 
responsible for the enhancement  of the couplings.
In Fig.~\ref{fig:astro_bounds_A} the parameter $\kappa$ has been fixed to its 
minimum value $\kappa =0$ which implies that $\gae$ is 
determined solely by the radiative contributions and hence 
strongly suppressed. 
Fig.~\ref{fig:astro_bounds_B} corresponds to 
to the maximum value allowed by  perturbative unitarity on the charm Yukawa 
coupling $\kappa=0.92$, in which case $\gae$ is exponentially enhanced. 


The hatched grey area in the figures represents the region of parameters excluded by the duration
of the SN1987A neutrino signal.
Historically, observation of the SN1987A neutrinos has provided 
the strongest bounds on the axion-nucleon couplings~\cite{Burrows:1988ah,Burrows:1990pk,Keil:1996ju,Carenza:2019pxu}.
In the plots we use a state-of-the-art determination of the SN1987A limit 
from Ref.~\cite{Carenza:2019pxu}. 
The lower edge of the grey region  corresponds to axions so weakly coupled
that, while  they can freely stream out of the supernova, the amount of energy they can carry  away does not shorten sufficiently the neutrino burst duration. 
In this regime,  the limit applies to the following combination 
of  couplings  $\sqrt{\gan^2 + 0.6 \gap^2+0.5g_{an}g_{ap}}$. 
The upper edge is determined by the onset of the trapping regime, in which 
axion interactions are sufficiently strong  that their mean free path 
is smaller than the proto-NS radius so that, like neutrinos, 
axions remain trapped inside the star for a sufficiently long time  not  
to affect significantly the 
neutrino burst duration~\cite{Burrows:1990pk,Carenza:2019pxu}. 
Both bounds are  afflicted by several uncertainties and should be 
taken with a grain of salt. However, 
the upper edge of the region is already 
covered by the CAST results (see Section~\ref{sec:axionexp}) 
so that determining the precise values of the couplings 
for the onset of the trapping regime is not crucial. 
In contrast, a reliable assessment of the validity of the 
limit for the free streaming regime is an 
important issue. 
In the pictures we use the bound derived in~\cite{Carenza:2019pxu}.
However, it was recently claimed  that axion-pion interactions may 
contribute  more than previously thought to the axion emissivity,  
in which case the limit could be sizeably  stronger~\cite{Carenza:2020cis}.

At low values of the axion mass,  the region in which 
axions can account for the M7 anomaly is in 
strong tension with several astrophysical observations. 
These include the recent NuSTAR search for hard X-rays emission from  Betelgeuse~\cite{Xiao:2020pra} and from the Quintuplet and Westerlund~1 super star clusters~\cite{Dessert:2020lil}.
Both analyses exclude the region $m_a\lesssim 10^{-10}\,$eV  for 
$g_{a\gamma}\gsim$  few$\times 10^{-12}\,{\rm GeV}^{-1}$.
A similar region is also excluded by the Fermi LAT bound on 
diffuse gamma rays  that would result from conversion of SN 
axions into photons in the Galactic magnetic
field~\cite{Calore:2020tjw}.
At even smaller masses, $m_a\lesssim 10^{-11}\,$eV, CHANDRA observations of NGC1275 set a slightly stronger bound on the axion-photon coupling~\cite{Reynolds:2019uqt}.
The strongest constraint is, however, implied by the non-observation 
of gamma rays by the Solar Maximum Mission (SMM) at the time of SN1987A explosion~\cite{Brockway:1996yr,Grifols:1996id,Payez:2014xsa}, 
and corresponds to the purple hatched region in the figures. 
We see from the pictures that this bound is especially relevant 
for $n\gsim 20$, and this is because  the  enhancement 
of the  axion-nucleon couplings in this regime strongly enhances  
the  SN axion emissivity.

The orange line labelled HB in the figures represents the horizontal branch (HB) star bound~\cite{Raffelt:1987yu,Ayala:2014pea}, which constraints the axion-photon coupling from the ratio of the  
observed number of stars in the HB and RGB 
in globular clusters (the bound shown in the figures is from the latest analysis in Ref.~\cite{Ayala:2014pea}).

The strongest astrophysical bound on the 
axion-electron coupling $\gae$ is derived from observations of 
red giant branch 
(RGB) stars~\cite{Straniero:2018fbv,Capozzi:2020cbu,Straniero:2020iyi}. 
The RGB bound in Fig.~\ref{fig:astro_bounds_A}
is very weak since  in model A,  $\kappa\approx 0$ implies 
$\gae\approx 0$ at tree level.
However, as can be seen from Fig.~\ref{fig:astro_bounds_B},  
$\gae$ gets exponentially enhanced for $\kappa=0.92$, and then the RGB bound dominates over all 
other limits, including that from SN1987A.
(The same can occur also in other constructions, as for example in model~B 
at large $\kappa$. In this specific case, however,  $\gaN$ also gets enhanced, 
and the RGB bound is only slightly more constraining than the SN1987A limit.)

\begin{figure}[t]
\centering
\includegraphics[width=8.7cm]{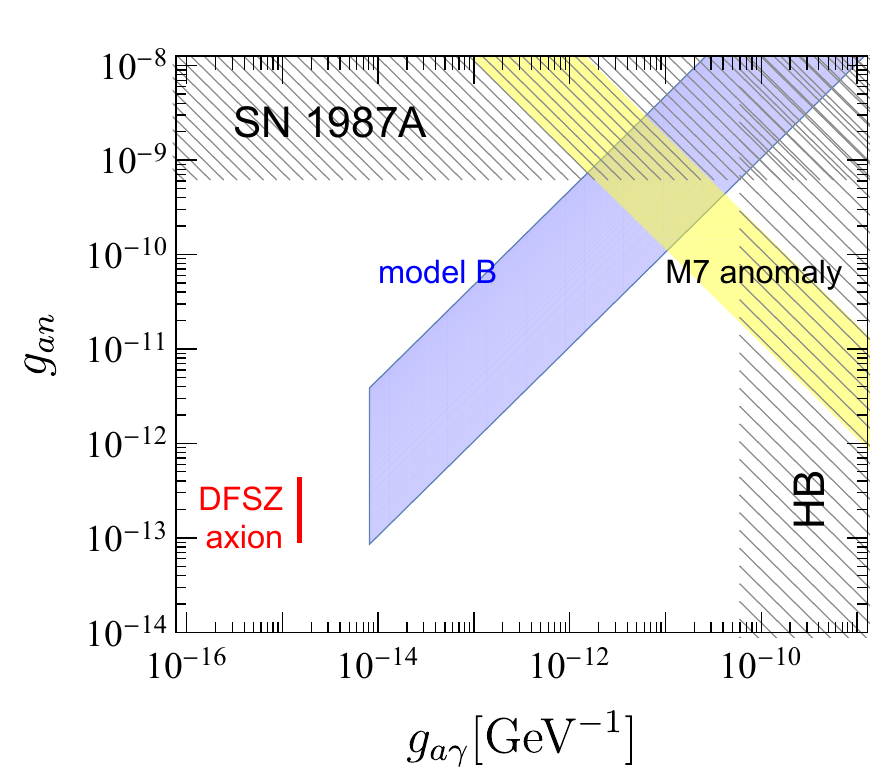}
\caption{
The $g_{an}$-$g_{a\gamma}$ parameter space for  $m_a = 10\,\mu$eV.
The vertical red segment corresponds to the DFSZ axion  
with $\gan$ within the phenomenologically allowed range. 
The blue strip corresponds to model B for $0\lesssim \kappa\lesssim 0.92$.
The grey hatched areas are   
excluded respectively by the SN1987A and HB bounds.
 The (2$\sigma$) region for which  
the M7 anomaly can be explained in terms of axion/ALP 
emission/conversion~\cite{Buschmann:2019pfp}  
corresponds to the yellow band. 
\label{fig:mass_10_meV}
}
\end{figure}

Finally, the yellow 
area in the figures corresponds to the hint from 
a $2\,\sigma$ fit to the observed  excess of hard X-ray 
events from the Magnificent Seven (M7) group of NS~\cite{Buschmann:2019pfp}.
The origin of this anomaly is not  understood, 
and it was speculated in Ref.~\cite{Buschmann:2019pfp} 
that the excess might be attributed to axion-like particles
(ALPs) with couplings to both photons and neutrons.  
The axion-neutron coupling would be responsible 
for the  production of ALPs in the hot NS core, 
while the coupling to photons would allow the 
ALP to be converted into photons 
in the strong magnetic fields surrounding the NS. 
The resulting photon flux would then be detected by X-ray detectors, 
such as XMM-Newton and Chandra, 
and would correspond to the excesses observed by these instruments.  
Notice that the observations demand quite large couplings 
and yet a small mass,  in order for the axions to be 
efficiently converted into photons in the magnetic field.
More specifically, Ref.~\cite{Buschmann:2019pfp} found that 
the mass should not exceed $\sim 20\,\mu$eV, while the couplings 
should satisfy the relation
$g_{a\gamma} \gan \sim $ a few $10^{-21}\,{\rm GeV^{-1}}$. 
Given the upper limit on the mass, such couplings are prohibitively large for 
canonical QCD axion models, such as DFSZ (see the red vertical segment in Fig.~\ref{fig:mass_10_meV}), and that is why  one would more 
generically invoke an  ALP.
However, our construction is  versatile enough to provide a QCD axion that fits well the data 
even in this extreme case. In fact, thanks to the exponential  
enhancement of the  axion couplings to nucleons and photons,  
axion production and their conversion into photons can proceed 
with  sufficient rates  even in the small mass window. 
This is shown in Fig.~\ref{fig:mass_10_meV}, where we 
present the axion parameter space for $m_a=10\,\mu$eV.
The red line refers to the DFSZ axion, constrained by 
perturbative unitarity. 
The blue band represents the parameter space for model B 
with $\kappa$ varying in the allowed range $0\lesssim \kappa\lesssim 0.92$ 
and $2\leq n\leq 40$.
We see that the photophilic and nucleophilic axion of model B  
can have sufficiently large couplings to account for the M7 anomaly 
even at this low mass value.  
As regards model A, the combination of the SN1987A and RGB bounds 
always excludes the  region of couplings hinted by the M7 anomaly. 
In fact, in order to evade the SN1987A bound, model A requires 
a large $\kappa$ which, in turn, is strongly constrained by the RGB bound. 

\subsection{Experimental bounds and perspectives}
\label{sec:axionexp}

\begin{figure*}[t]
\centering
\includegraphics[width=12cm]{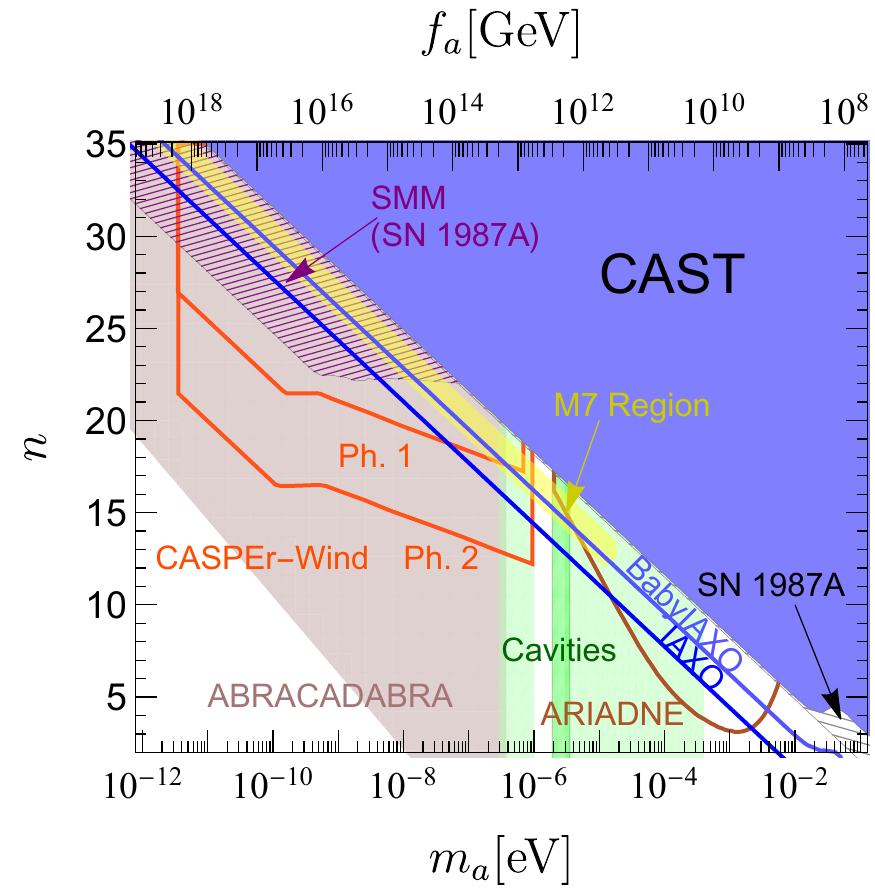}
\caption{
Current bounds and expected sensitivities of the next generation of axion probes
confronted with model B with $\kappa \approx 0$.
In green are the cavity experiments, with ADMX in darker green.
The other green areas comprise the projected sensitivities of KLASH~\cite{Alesini:2019nzq}, 
CAPP~\cite{Semertzidis:2019gkj}, ORGAN~\cite{McAllister:2017lkb}, and MADMAX~\cite{Brun:2019lyf}.
The region in light purple indicates the reach of ABRACADABRA 
phase 1 with resonant signal readout~\cite{Kahn:2016aff}. 
The sensitivity of CASPEr-Wind  phase 1 and 2 
to $\gaN$~\cite{JacksonKimball:2017elr} corresponds to the two red lines. 
The reach of ARIADNE is that enclosed in the brown line, at relatively large 
masses~\cite{Arvanitaki:2014dfa}, and the 
projected  sensitivities of IAXO~\cite{Armengaud:2019uso} and BabyIAXO~\cite{Abeln:2020ywv}
are given by the two blue lines.
\label{fig:future_experiments}
}

\end{figure*}

In this section we discuss the potential of current and next generation 
axion experiments to probe the nucleophilic models.
For definiteness, we show the results for model~B, 
though many of our conclusions apply
to model~A as well. 

Current axion searches are probing (mostly) the axion coupling to photons.
However, some proposals for future search strategies suggest exploiting 
also the couplings to nucleons, and in particular to neutrons
(see~\cite{Irastorza:2017,Sikivie:2020zpn} for comprehensive reviews).
Presently, one of the tightest  experimental bounds 
come from the  CERN Axion Solar Telescope (CAST), 
which has probed the axion-photon coupling down to
$g_{a\gamma}=0.66\times 10^{-10}\,$GeV$^{-1}$ for masses up to $0.02\,$eV~\cite{Anastassopoulos:2017ftl}.

In Fig.~\ref{fig:future_experiments}
we show the viable and the excluded regions for model~B 
for $\kappa\approx 0$. 
The region excluded by CAST is shown in blue.
The  region 
probed by the cavity haloscope ADMX~\cite{Du:2018uak,Braine:2019fqb}, 
which covers so far only a narrow mass window,  is shown in dark green. 
Next generation  cavity experiments are expected to  
probe a considerable wider mass range from $\sim 300\,\mu$eV to $\sim 0.5\,$meV,
as shown by the light green region. 
The region includes the KLASH~\cite{Alesini:2019nzq}, 
CAPP~\cite{Semertzidis:2019gkj},  ORGAN~\cite{McAllister:2017lkb}, and  MADMAX~\cite{Brun:2019lyf} 
proposals.\footnote{The gap between $m_a=1\,\mu$eV and 2$\,\mu$eV is also target by new haloscope proposals. 
The  UPLOAD-DOWNLOAD experiment~\cite{Thomson:2019aht} might close it 
in the next few years, though it is currently  exploring 
a lower mass region.}
The region at low masses can be probed by  
ABRACADABRA~\cite{Kahn:2016aff} and CASPEr-Wind~\cite{JacksonKimball:2017elr}.
As shown in the figure, these experiments are expected to cover a large portion of the 
sub $\mu$eV region of the nucleophilic axion model. Finally, the ARIADNE proposal~\cite{Arvanitaki:2014dfa} would allow to probe 
 the large mass region through the axion coupling to neutrons. 
 However, the corresponding limits will hold only 
 under the assumption that the  amount of CP violation
 is maximal and saturates the neutron electric dipole moment bound. 

The next generation of solar axion searches  
will have a considerable higher sensitivity than CAST, and 
will be able to set limits for a wide range of axion masses. 
The International Axion Observatory (IAXO)~\cite{Armengaud:2019uso}, 
will have enough sensitivity to test the axion explanation 
of the M7 anomaly anomaly. 
BabyIAXO~\cite{Abeln:2020ywv}, an intermediate experimental stage of IAXO which is 
expected to become operative by the mid of the current decade,  
will already expand considerably the region probed by CAST, and 
can already probe a large part of the M7 region.
Finally, the forthcoming light-shining-through-walls experiment 
ALPS~II~\cite{Bahre:2013ywa} (not shown in Fig.~\ref{fig:future_experiments}), 
which is expected to take data starting from 2021, 
will also surpass CAST, probing the parameter space almost to the level 
of sensitivity of BabyIAXO, although in a smaller mass window 
$m_a\lesssim 0.1\,$meV.

Let us note that, with the exception of
CASPEr-Wind and ARIADNE, all the axion experiments included in 
 Fig.~\ref{fig:future_experiments} probe the axion-photon
coupling, which has the same form in model~A than in model-B, and  
in particular it does not depend on $\kappa$.
Hence, similar results can be expected for model~A and for 
other choices of $\kappa$.

\subsection{Cosmology} 
\label{sec:cosmo}

Exponentially enhanced axion couplings may lead  in the primordial Universe to a thermal population of hot axions, and this would modify the  effective number of neutrinos with respect to  
that inferred by the Planck collaboration  from CMB measurements~\cite{Aghanim:2018eyx}. 
Given that the axion couples strongly to first generation quarks, we will first consider the regime in which axions decouple from the thermal bath at 
$T_{\rm d} \lesssim 1\,$GeV. In this regime  quarks are bounded into nucleons 
whose number density is Boltzmann suppressed, and into pions that 
(for $T_{\rm dec}\gsim m_\pi$) are as abundant as  photons, so that 
axion coupling to pions is the relevant quantity.
The axion thermal production rate $\Gamma_{a\pi}$ has been estimated in~\cite{Hannestad:2005df}:
\begin{align}
\label{eq:Gapi}
\Gamma_{a\pi} \simeq 0.215 C_{a \pi}^2 \frac{T^5}{f_a^2 f_\pi^2} h
\left( \frac{m_\pi}{T}\right) \ ,
\end{align}
where $f_\pi = 93$ MeV, $h$ is an exponentially 
decreasing function satisfying with $h(0)=1$ and is tabulated in~\cite{Hannestad:2005df}, 
and the axion-pion coupling is:
\begin{align}
C_{a \pi} &= - \frac{1}{3} \left(  c^0_{u} - c^0_{d} - \frac{m_d - m_u}{m_u+m_d}\right)
\, .
\end{align} 
Taking as an example model A, and neglecting the model-independent contribution 
(the third term in parenthesis) we have: $C_{a\pi}\approx   - {2 c^0_{u} }/{3}$. 
Using this coupling  together with   \eqn{eq:Gapi},
and recalling that the decoupling temperature is defined by the condition 
$ \Gamma_{a\pi} (T_{\rm d}) \simeq H (T_{\rm d})$
where $H(T) = 1.66 \sqrt{g_{\rm eff}}\; T^2/m_{\rm P}$ is the Hubble parameter,
$m_{\rm P}$ the Planck mass and $g_{\rm eff}$ the effective number of 
relativistic degrees of freedom in thermal equilibrium,  we can straightforwardly 
derive  the  contribution  $\Delta N_{\rm eff} = N_{\rm eff}  - N^{\rm SM}_{\rm eff}$ due 
to a thermal axion population  
\begin{align}
\Delta N_{\rm eff}  &\simeq  
\frac{13.6}{ g_{\rm eff}^{4/3} (T_{\rm d})}\lesssim 0.36 \
\Rightarrow \   f_a^\mathrm{A} \gtrsim  4\cdot 10 ^7  c^0_{u}\,\mathrm{GeV} .
\end{align}
The  first numerical bound corresponds to the  $2\sigma$ 
 measurement of $N_{\rm eff}$ from the Planck collaboration~\cite{Aghanim:2018eyx}, while 
 the lower limit on the axion decay constant for model A  
 ($f_a^{\mathrm{A}}$) has been computed in the following way: 
 the Planck upper bound on $\Delta N_{\rm eff}$ implies 
 $g_{\rm eff} (T_\mathrm{d})\gsim 15.3$ which corresponds to a limit 
 on the decoupling  temperature $T_\mathrm{d} \gsim 66\,$MeV 
 (see Fig.1 in Ref.~\cite{Srednicki:1988ce}). 
 For this temperature the Hubble parameter is $H(T_\mathrm{d})\simeq 2.4\cdot 10^{-18}\,$MeV, and 
 the limit is then obtained from the out-of-equilibrium condition~\eqn{eq:Gapi}
using $C_{a\pi}^A=- 2 c^0_u/3$. Note that this limit is sub-dominant compared to the SN1987A bound on the 
nucleon couplings.

In the case where couplings to second and third generation are enhanced, thermalisation 
can occur via the $q g \to q a$ and $q q \to g a$ processes~\cite{Ferreira:2018vjj}. However, the predicted value of $\Delta N_{\rm eff}$ will only be in reach of future CMB-S4 experiments, leaving to a distant future the possibility of deriving 
constraints on the axion decay constant by using these processes. 

\section{Conclusions} 
\label{sec:concl}

We have presented in this work a simple structure to obtain a nucleophilic 
QCD axion. The key ingredient
of this construction is the presence of flavour-dependent Yukawa interactions of Higgs doublets with the SM quarks. In particular, one Higgs doublet with a very large PQ charge must interact with both the up and down quarks, leading to a cancellation of its contribution to the QCD anomaly 
and, at the same time, a large axion-nucleon coupling. We have further constructed an explicit realization of this scenario via a clockwork mechanism directly at the level of the Higgs doublets. 

Interestingly, we have shown that this construction can in fact generate an exponential enhancements of almost any axion coupling (barring the nEDM one), 
thus realising a scenario in which  the parameter region of 
QCD axion models can be extended to overlap with mass/couplings regions 
that are generally considered viable only in specific ALP 
models.\footnote{Clockwork constructions can 
find applications also beyond the QCD axion, as for example 
for ultralight ($m\sim 10^{-22}\,$eV) scalar DM, also 
called fuzzy DM~\cite{Hui:2016ltb}. 
In fact, breaking the parametric relationship between 
the mass and the couplings  might open up the possibility of direct 
detection also for ultralight pseudoscalars~\cite{Dror:2020zru}.
}
The two main restrictions of this setup is that one can only enhance the axion couplings with respect to the standard QCD axion cases, and that the model typically predicts flavour-dependent 
(as well as  flavour violating) interactions. 
Additionally, due to the gradation of the doublets PQ charges in steps of $2^i$ the model provides strong constraints on the allowed Yukawa couplings, similar to those found in simpler two Higgs doublet (2HD) models (see \app{sec:flavour}). It would be interesting to examine in more details to which point our clockwork inspired  multi-Higgs doublet model departs from the results found in these 2HD setups.

We have then analysed the phenomenology of two such nucleophilic models. First we considered a simple setup (model A), where the axion couplings to first generation quarks are strongly enhanced, along with the axion-photon interaction. We then studied a variation (model B) where the couplings to second generation quarks are instead boosted. The nucleon-axion couplings are then mostly generated by the axion interaction with the sea quarks in the nucleon.  We  emphasise 
that both scenarios can be easily tested in the near future by a large number of experiments, such 
as CASPEr-Wind, ABRACADABRA, or ARIADNE. In particular, the proposed model B provides an elegant solution for the excess of X-ray events originating from the ``Magnificent Seven'', a group of isolated NS, 
thanks to the light mass of the axion and its enhanced couplings to both photons and nucleons. We have further shown that such explanation will be probed by various upcoming axion experiments in the near future.

\section*{Acknowledgments}
\noindent
 L.D. and E.N.~are supported in part by the INFN ``Iniziativa Specifica'' Theoretical Astroparticle Physics (TAsP-LNF). 
The work of L.D.L.~is supported by the Marie Sk\l{}odowska-Curie Individual Fellowship  grant 
AXIONRUSH (GA 840791) 
and the Deutsche Forschungsgemeinschaft under Germany's Excellence Strategy 
- EXC 2121 Quantum Universe - 390833306.


\appendix

\addcontentsline{toc}{section}{Appendices}

\section{Photophilic and electrophilic axions}
\label{sec:philic}
As anticipated in Section \ref{sec:gaN_exp_idea}, 
the same structure put forth to give rise to a nucleophilic axion  
 can be straightforwardly adapted to generate an electrophilic or a photophilic axion. 
A dominantly photophilic axion can be obtained by coupling $H_n$ to the $\tau$ lepton only. 
In this case, the couplings become (model C):
\begin{align}
\label{eq:gagC}
g_{a\gamma}&\simeq - 2^n  \,  \alpha/({2\pi f_a})\,, \\ 
    \label{eq:gapC}
g_{a p} &\simeq  - \left(1.30 \,\kappa\, 2^{n-1} - \delta c_p^0 \right) \,  (m_p/f_a)\,, \\
\label{eq:ganC}
\gan &\simeq \left(1.24\,\kappa \,  2^{n-1}+\delta c_n^0 \right)  (m_n/f_a) \,, \\
\label{eq:gaeC}
g_{ae} &= \left(\kappa\, 2^{n-1} + \delta c^0_\ell\right) \, (m_e/f_a)\,, 
\end{align}
with, in the large $n$ limit,   
$8\cdot 10^{-6} \lesssim \kappa\lesssim 0.92$ where the 
 lower bound  is implied by perturbativity of the tau Yukawa coupling. 
Since in  this model all the fermion couplings are proportional to $\kappa$, 
in  the limit of small $\kappa$ the radiative contributions to $\gae\,,\gap$ and  $\gan$  
can became important, so they have been explicitly included  
in \eqs{eq:gapC}{eq:gaeC}.
For the electrons $\delta c_\ell^0$ is given in Eq.~\eqref{eq:c0e}. For the nucleons 
the same expression holds with the replacements $Q^2_\ell \to 0.88\, Q_u^2 - 0.39\, Q_d^2 =0.43$
for $\delta c^0_p$ and $Q^2_\ell \to 0.88\, Q_d^2 - 0.39\, Q_u^2 = -0.076$
for $\delta c_n^0$. 

A dominantly  electrophilic axion, can be obtained by coupling the electron 
to $\tilde H_n$, and the $b$ and $c$ quarks to respectively $\tilde H_n$ and  $H_n$ 
(the $t$ quark would then couple to $H_1$ and the $s$ to $\tilde{H}_0$, mimicking the structure in Eq.~\eqref{eq:gaNexpHL}). 
This Yukawa structure produces the following couplings (model D):
\begin{align}
\label{eq:gagD}
g_{a\gamma}&\simeq  \left(\frac{8}{3}-1.92\right) \,  \alpha/({2\pi f_a})\approx 0\,, \\ 
    \label{eq:gapD}
g_{a p} &\simeq   -2^{n-1}\times 0.003\,   (1+435\kappa) \,  (m_p/f_a)\,, \\
\label{eq:ganD}
\gan &\simeq  - 2^{n-1}\times  0.003\, (1-413\kappa) \,  (m_n/f_a) \,, \\
\label{eq:gaeD}
g_{ae} &= 2^{n-1}\, \left(\kappa-1\right)  \, (m_e/f_a)\,, 
\end{align}
The allowed range for the parameter $\kappa$ for this case is 
$2 \cdot 10^{-5}\lesssim \kappa \lesssim 0.92$ where the lower limit 
comes from the fact that the bottom quark couples to $H_n$.
We see that the axion is dominantly coupled to the electron, 
while the coupling to the proton and the neutron, which is mainly due to the 
bottom and charm sea quarks, is clearly subdominant. 
Moreover, in the approximation in which only the exponentially enhanced terms 
are kept, the axion is decoupled from the photon, and recalling 
that $\gag$ does not receive corrections proportional to the  fermion couplings
(see the discussion in Section~\ref{sec:gaN_exp_idea})  this result holds at all orders. 
Finally,  let us note that by replacing  in  models 
A, B, C and D the doublet $H_1$ by $H_i$ with $i\gtrsim1$, the coefficient of the QCD anomaly is  enhanced as $2N \sim 2^{i-1} \dchi$. This can be used to suppress the model-dependent contribution of some selected axion couplings.

\section{Multiple Higgses and Yukawa textures}
\label{sec:flavour}

Generation dependent $U(1)$ symmetries acting on the quark and Higgs fields 
can provide a powerful tool to reduce the number of fundamental flavour 
parameters. The issue of which $U(1)$ symmetries can enforce the maximum  parameter 
reduction consistently with experimental data (six quark masses, three mixing angles and 
one non-vanishing CP phase) in a scenario with  two Higgs doublets (2HD) was systematically 
addressed in Ref.~\cite{Bjorkeroth:2018ipq} (see also Ref.~\cite{Bjorkeroth:2019ndr} 
for an analogous study in the lepton sector equipped within the type-I seesaw).   
The interest in such constructions is that Yukawa structures with texture zeros 
are generally more predictive, and it would  be certainly more satisfactory   
if the vanishing of some Yukawa entry could be justified in terms of some symmetry.
The construction discussed in this paper is characterised by a proliferation of Higgs doublets, 
each one with its own PQ charge, and it is reasonable to ask if in this case 
a  $U(1)$ PQ symmetry could still be effective 
to enforce texture zeros in the quark Yukawa matrices.

Let us recall how parameter reduction proceeds as a consequence 
of a $U(1)$ symmetry. Consider for example the Yukawa sector for  up-type 
quarks  of the second and third generation. With  sub-indices referring to  
generations, e.g.~$\mX_{32} \equiv \mX(\bar t_{L} c_{R}) $ 
it is easy to see that the charges of quark bilinears  must satisfy 
the following identity: $\mX_{23}+\mX_{32} = \mX_{33}+\mX_{22}$. 
This can be translated into a relation for the Higgs charges. 
Referring for example to the couplings of model~A in \eqn{eq:gaNexpHL} 
we have $\mX_{22}=-\mX_0$ and $\mX_{33}=-\mX_1$,  
and choosing $Y_{23}$ to be non-vanishing (e.g.~to generate 
a particular CKM mixing) means that $\mX_{23}$ must match the charge of one  Higgs doublet, e.g.~$\mX_{23}= - \mX_j$.
$Y_{32}$ would then be a texture zero  unless $\mX_{32}$ also matches the charge 
of some Higgs $H_k$, that is:    
\begin{equation}
\label{eq:tmX}
\mX_j + \mX_k = \mX_0 + \mX_1=\dchi + 2 \chi_0.
\end{equation}
Thus $(\mX_j,\mX_k)= (\mX_0,\mX_1)$ or $(\mX_1,\mX_0)$  
allows for both $Y_{23},Y_{32}\neq 0$. Only two Higgs doublets are involved, and in fact 
this is  the same solution one has in the 2HD case.
For  $j,k\geq 1$, we see from the structure of the charges given in \eqn{eq:Xiter} 
that  the coefficient of $\dchi$ 
 can never be matched since $2^{j-1}+2^{k-1}=1$ has no solution. Hence 
in the clockwork inspired  multi-Higgs doublet scenario there are no 
additional possibilities with respect to the 2HD case to avoid one zero 
texture in the (23) Yukawa submatrix. 
Repeating this argument for the (13)  submatrix one would obtain 
$2^{j-1}+2^{k-1}=2^{n-1}+1$ which has again the same solutions 
$(j,k)=(1,n)$ or $(n,1)$ in terms of just two Higgs doublets than the 2HD case. 
Only for the (12) submatrix, for which the  $j,k\geq 1$ condition reads 
$2^{j-1}+2^{k-1}=2^{n-1}$,  the solution $j=k=n-1$ opens up a new possibility 
for avoiding zero textures in terms of three Higgs doublets $\{H_0,H_{n-1},H_n\}$.  Clearly,  the same considerations will also hold for the other possible PQ charge assignments leading to model B, C and D.

We conclude that, in spite of the presence of a large number of Higgs doublets 
all with different charges, thanks to the hierarchical distribution of the charge values the $U(1)_{\mathrm{PQ}}$ symmetry of the 
multi-Higgs clockwork model remains  well suited to enforce zero textures in the 
Yukawa matrices of the quarks. We can therefore expect that most of the results 
derived in Refs.~\cite{Bjorkeroth:2018ipq,Bjorkeroth:2019ndr} for 
the 2HD scenario will  hold also in the present case.  

\bibliography{bibliography}

\clearpage

\end{document}